\documentclass[%
superscriptaddress,
preprint,
showpacs,preprintnumbers,
 amsmath,amssymb,
 aps,
]{revtex4-1}

\usepackage{graphicx}
\usepackage{dcolumn}
\usepackage{bm}

\usepackage{url}

\begin{document}

\preprint{APS/123-QED}

\title{A collective opinion formation model under Bayesian updating and confirmation bias}

\author{Ryosuke Nishi}
\affiliation{
National Institute of Informatics, 2-1-2 Hitotsubashi, Chiyoda-ku, Tokyo 101-8430, Japan
}%
\affiliation{
JST, ERATO, Kawarabayashi Large Graph Project, 2-1-2 Hitotsubashi, Chiyoda-ku, Tokyo 101-8430, Japan
}%
\affiliation{
Department of Mathematical Informatics,The University of Tokyo, 7-3-1 Hongo, Bunkyo-ku, Tokyo 113-8656, Japan
}%
\author{Naoki Masuda}%
 \email{masuda@mist.i.u-tokyo.ac.jp}
\affiliation{%
Department of Mathematical Informatics,The University of Tokyo, 7-3-1 Hongo, Bunkyo-ku, Tokyo 113-8656, Japan
}%

\date{\today}

\begin{abstract}
We propose a collective opinion formation model with a so-called confirmation bias. The confirmation bias is a psychological effect with which, in the context of opinion formation, an individual in favor of an opinion is prone to misperceive new incoming information as supporting the current belief of the individual. Our model modifies a Bayesian decision-making model for single individuals [M. Rabin and J. L. Schrag, Q. J. Econ. \textbf{114}, 37 (1999)] for the case of a well-mixed population of interacting individuals in the absence of the external input. 
We numerically simulate the model to show that all the agents eventually agree on one of the two opinions only when the confirmation bias is weak. Otherwise, the stochastic population dynamics ends up creating a disagreement configuration (also called polarization), particularly for large system sizes. A strong confirmation bias allows various final disagreement configurations with different fractions of the individuals in favor of the opposite opinions.
\begin{description}
\item[PACS numbers] 87.23.Ge, 02.50.Ey, 02.50.Le 
\end{description}
\end{abstract}

\maketitle


\section{Introduction} \label{sec:Introduction}
There are various models of collective opinion formation in which agents modify their opinions according to interaction with other agents \cite{CastellanoFortunatoLoreto2009RevModPhys,KrapivskyRednerBen-Naim2010AKineticView}. Opinion formation is a dynamic process: for example, interaction between agents makes their opinions approach each other. An important problem in opinion dynamics is to examine when an agreement (i.e., consensus) among all the agents occurs. Complete agreement is rarely observed in the real world \cite{Kuran1995PrivateTruthsPublicLies,HuckfeldtJohnsonSprague2004PoliticalDisagreement}. However, it is an established fact that opinion dynamics under the voter model, a classical opinion model in statistical physics and probability theory, inevitably reaches agreement in finite populations \cite{CastellanoFortunatoLoreto2009RevModPhys,DonnellyWelsh1983MathProcCambPhilSoc,SoodRedner2005PhysRevLett,AntalRednerSood2006PhysRevLett,SoodAntalRedner2008PhysRevE,MasudaOhtsuki2009NewJPhys}. The majority rule model has a similar feature \cite{CastellanoFortunatoLoreto2009RevModPhys,Galam2002EurPhysJB,ChenRedner2005PhysRevE}. 
Partly motivated by this discrepancy, various extensions of voter and majority rule models and different models of collective opinion formation have been proposed to account for the disagreement in finite populations. Examples include 
the Deffuant model \cite{DeffuantNeauAmblardWeisbuch2000AdvComplexSyst}, 
language competition models 
\cite{PatriarcaLeppanen2004PhysicaA,PatriarcaHeinsalu2009PhysicaA,MiraParedes2005EurophysLett,MiraSeoaneNieto2011NewJPhys,CastelloEguiluzSanMiguel2006NewJPhys,VazquezCastelloSanMiguel2010JStatMech,ChapelCastelloBernardDeffuantEguiluzMartinMiguel2010PLoSONE,FujieAiharaMasuda2012JStatPhys}, 
voter-like models on adaptive networks \cite{Vazquez2008,Kozma2008,Holme2006,Nardini2008}, voter model under partisan bias (the assumption that agents naturally prefer one opinion) \cite{MasudaGibertRedner2010PRE,MasudaRedner2011JStatMech}, and variations of Axelrod's cultural dynamics (see \cite{CastellanoFortunatoLoreto2009RevModPhys} for references). Theoretical models have also been proposed in social sciences to explain disagreement in the context of polarization. For example, prior beliefs or initially received signals can cause disagreement between agents, even if they receive the same public signals from then on \cite{DixitWeibull2007PNAS,ZimperLudwig2009JRiskUncertain,AcemogluVictorMuhamet2009Fragility,AcemogluOzdaglar2011DynGamesAppl,AndreoniMylovanov2012AmEconJ-Microecon}. 

Although there is a plethora of studies addressing the problem of agreement and disagreement in opinion dynamics, we propose a model incorporating two factors that are relevant to human behavior: Bayesian belief updating and confirmation bias.
Bayesian belief updating is commonly used in studies of the decision making of agents receiving uncertain information 
\cite{Binmore2008RationalDecisions,AcemogluOzdaglar2011DynGamesAppl,Martins2009JStatMech}. 
The confirmation bias is a psychological bias inherent in humans, in which an agent inclined towards an opinion tends to misperceive incoming signals as supporting the agent's belief \cite{Plous1993ThePsychologyofJudgmentandDecisionMaking,Nickerson1998RevGenPsychol}. 
A non-Bayesian model with the confirmation bias was previously proposed for explaining the influences of media and interactions between agents \cite{DeffuantHuet2009COMPLEXITY}.

We are not the first to study opinion formation under the Bayesian updating and confirmation bias. In the framework of single agent opinion formation, Rabin and Schrag showed that the confirmation bias triggers overconfidence and can cause the individual to hold incorrect beliefs, even if it receives a series of external signals suggesting the true state of the world \cite{RabinSchrag1999QJECON}. Orl\'{e}an studied the Bayesian dynamics of agents subjected to the confirmation bias, interacting through the mean field \cite{Orlean1995JEconBehavOrgan}. The model yields agreement or disagreement depending on the parameter values.

In this study, motivated by the Rabin-Schrag model \cite{RabinSchrag1999QJECON}, we propose a model of collective opinion formation with a confirmation bias. We model direct peer-to-peer interactions between agents (not through the mean field) and their effects on the Bayesian updating of each agent. To study the pure effects of interactions among agents, we do not assume that agents receive signals from the environment as in previous studies \cite{AndreoniMylovanov2012AmEconJ-Microecon}. We numerically simulate the model to reveal the conditions under which the populations of agents agree and disagree, depending on the values of parameters such as the strength of the confirmation bias, fidelity of the signal, and the system size. 

\section{Model} \label{sec:Model}
Our model modifies the Bayesian decision-making model proposed by Rabin and Schrag \cite{RabinSchrag1999QJECON} in two main ways. First, we consider a well-mixed population of Bayesian agents that interact with each other; Rabin and Schrag focused on the case of the single agent. Second, agents do not receive external signals from the environment in our model. In the Rabin-Schrag model, such an external signal, which represents the ``correct" answer in the binary choice situation (i.e., the true state of nature), is assumed. By making the two changes, we concentrate on collective opinion formation by Bayesian agents, whereby there are two possible alternative opinions of equal attractiveness.

We label the $N$ agents $1,\ldots,N$ and denote the opinion of agent $i$ ($i=1,\ldots,N$) by $x_i \in \left\{ {\rm A}, {\rm B} \right\}$, where A and B are the alternative opinions. We assume that agents are not perfectly confident in their opinions. To model this factor, we adopt the Bayesian formalism used by Rabin and Schrag \cite{RabinSchrag1999QJECON}. We denote by $\Pr(x_i={\rm A})$ the strength of the belief (hereafter, simply the belief) with which agent $i$ believes in opinion A. A parallel definition is applied to $\Pr(x_i={\rm B})$. It should be noted that $\Pr(x_i={\rm A}) \ge 0$, $\Pr(x_i={\rm B}) \ge 0$, and $\Pr(x_i={\rm A}) + \Pr(x_i={\rm B}) = 1$. If $\Pr(x_i={\rm A}) = 1/2$, agent $i$ is indifferent to either opinion. 

We update the agent's belief as follows. 
The time $t$ starts from $t = 0$. Upon every updating of an agent's belief, we add $1/N$ to $t$ such that the belief of each agent is updated once per time unit on average. In an updating event, 
we select an agent $i$ to be updated with equal probability $1/N$. Agent $i$ refers to agent $j$'s opinion for updating $i$'s belief $\Pr(x_i={\rm A})$, where $j$ ($\neq i$) is selected with equal probability $1/(N-1)$ from the population. Agent $j$ imparts a signal $s \in \left\{a, b \right\}$, where $a$ and $b$ correspond to $j$'s opinions A and B, respectively. We assume that the probabilities that agent $j$ imparts $s=a$ and $s=b$ are given by
\begin{align} 
\Pr(s=a) &= \Pr(s=a | x_j={\rm A}) \Pr(x_j={\rm A}) + \Pr(s=a | x_j={\rm B}) \Pr(x_j={\rm B}) \nonumber \\
&= \theta \Pr(x_j={\rm A}) + (1 - \theta) \Pr(x_j={\rm B}) \nonumber \\
&= 1 - \theta + (2 \theta - 1) \Pr(x_j={\rm A})
\label{eq:Pja}
\end{align}
and
\begin{align} 
\Pr(s=b) &= \Pr(s=b | x_j={\rm A}) \Pr(x_j={\rm A}) + \Pr(s=b | x_j={\rm B}) \Pr(x_j={\rm B}) \nonumber \\
&= (1 - \theta) \Pr(x_j={\rm A}) + \theta \Pr(x_j={\rm B}) \nonumber \\
&= \theta - (2 \theta - 1) \Pr(x_j={\rm A}),
\label{eq:Pjb}
\end{align}
respectively, where
\begin{align} 
\theta=\Pr(s=a|x_j={\rm A})=\Pr(s=b|x_j={\rm B})
\label{eq:theta}
\end{align}
represents the reliability of the signal, and $1/2 \le \theta < 1$. If $j$ is confident in its own opinion and the transformation from $j$'s belief [i.e., $\Pr(x_j={\rm A})$] to $j$'s output signal (i.e., $a$ or $b$) is reliable, signals $a$ and $b$ are likely to indicate opinions A and B, respectively. In the limit $\theta \to 1$, $\Pr(s=a) \approx \Pr(x_j={\rm A})$ and $\Pr(s=b) \approx \Pr(x_j={\rm B})$. If $\theta=1/2$, $\Pr(s=a)=\Pr(s=b)=1/2$ such that $s$ does not convey any information about $j$'s belief. We implicitly assume that all the agents share the same value of $\theta$ and that they know this fact when performing the Bayesian update, as described below.

\begin{figure}[t]
\includegraphics[width=1.0\linewidth]{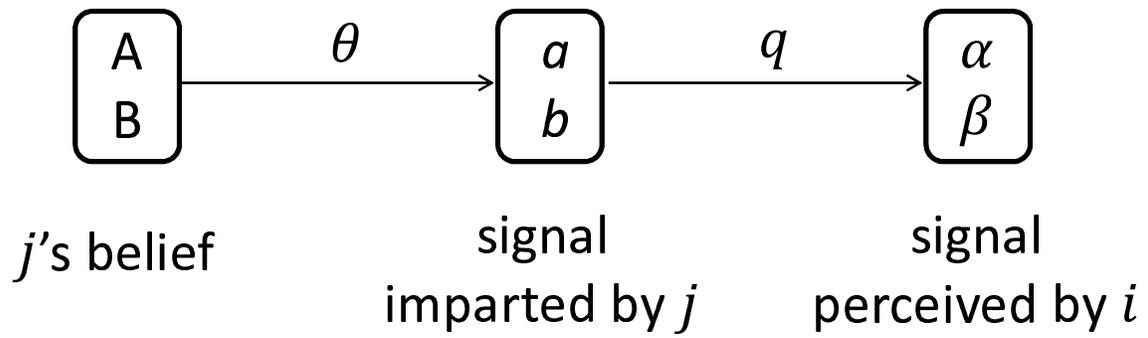}
\caption{
Schematic of signal conversion. Signals $a$ and $\alpha$ correspond to A. Signals $b$ and $\beta$ correspond to B. $\theta$ is the reliability of the signal, and $q$ is the strength of the confirmation bias. 
}
\label{fig:signals}
\end{figure}

When agent $j$ imparts signal $s \in \left\{ a,b \right\}$, agent $i$ is assumed to perceive a subject signal $\sigma \in \left\{ \alpha,\beta \right\}$, where $\alpha$ and $\beta$ correspond to ${\rm A}$ and ${\rm B}$, respectively. The flow of the signal conversion is depicted in Fig. \ref{fig:signals}. If agent $i$ is not subject to the confirmation bias, $\alpha$ and $\beta$ are equal to $a$ and $b$, respectively. Otherwise, agent $i$ may misinterpret the signal imparted by agent $j$, depending on the prior exposure of agent $i$ to other signals. Following Rabin and Schrag \cite{RabinSchrag1999QJECON}, we define
\begin{align} 
\Pr[\sigma=\alpha | s=a, \Pr(x_i={\rm A}) \ge 1/2] = \Pr[\sigma=\beta  | s=b,\Pr(x_i={\rm A}) \le 1/2] = 1
\label{eq:s_to_sigma_bias_1}
\end{align}
and
\begin{align} 
\Pr[\sigma=\alpha | s=b,\Pr(x_i={\rm A}) > 1/2] = \Pr[\sigma=\beta  | s=a,\Pr(x_i={\rm A}) < 1/2] = q, 
\label{eq:s_to_sigma_bias_2}
\end{align}
where $q$ ($0 \le q \le 1$) parameterizes the strength of the confirmation bias. Equation \eqref{eq:s_to_sigma_bias_2} states that an agent preferring opinion A misinterprets an arriving $b$ signal as ${\rm A}$ (i.e., $\sigma=\alpha$) with probability $q$. If $q=0$, the confirmation bias is absent, and $s=a$ and $s=b$ are always converted to $\sigma=\alpha$ and $\sigma=\beta$, respectively. If $q=1$, the agent perceives the signal that is consistent with its current preference [i.e., $\alpha$ if $\Pr(x_i={\rm A})>1/2$ and $\beta$ if $\Pr(x_i={\rm A})<1/2$], irrespective of the signal imparted by agent $j$ (i.e., $a$ or $b$). The other conditional probabilities can be readily derived from Eqs. (\ref{eq:s_to_sigma_bias_1}) and (\ref{eq:s_to_sigma_bias_2}). For example, Eq. (\ref{eq:s_to_sigma_bias_1}) implies 
\begin{align} 
\Pr[\sigma=\beta|s=a, \Pr(x_i={\rm A})>1/2] = 1 - \Pr[\sigma=\alpha|s=a, \Pr(x_i={\rm A})>1/2] = 0, 
\label{eq:s_to_sigma_bias_3}
\end{align}
and Eq. (\ref{eq:s_to_sigma_bias_2}) implies 
\begin{align} 
\Pr[\sigma=\alpha|s=a, \Pr(x_i={\rm A})<1/2] = 1- \Pr[\sigma=\beta|s=a, \Pr(x_i={\rm A})<1/2] = 1-q.
\label{eq:s_to_sigma_bias_4}
\end{align}

Then, by using the Bayes' theorem, we update agent $i$'s belief $\Pr(x_i={\rm A} | \sigma)$ on the basis of the old belief $\Pr(x_i={\rm A})$ [$=1-\Pr(x_i={\rm B})$] and the perceived signal (i.e., $\alpha$ or $\beta$). The perceived signal may be different from the received signal (i.e., $a$ or $b$) because of their confirmation bias [Eqs. \eqref{eq:s_to_sigma_bias_1} and \eqref{eq:s_to_sigma_bias_2}]. We assume that agents are not aware that they may be subject to the confirmation bias. Agents use the subjective conditional probabilities given by
\begin{align} 
\overline{\Pr}(\sigma=\alpha | s=a) = \overline{\Pr}(\sigma=\beta | s=b) = 1
\label{eq:eq:s_to_sigma_nobias_1of2}
\end{align}
and  
\begin{align}
\overline{\Pr}(\sigma=\alpha | s=b) = \overline{\Pr}(\sigma=\beta | s=a) = 0
\label{eq:eq:s_to_sigma_nobias_2of2}
\end{align} 
to perform the Bayesian update. The posterior belief $\Pr(x_i={\rm A} | \sigma)$ is given by  
\begin{align} 
&\Pr (x_i={\rm A} | \sigma)
= \dfrac{\Pr(\sigma | x_i={\rm A}) \Pr(x_i={\rm A})}{\Pr(\sigma | x_i={\rm A}) \Pr(x_i={\rm A}) + \Pr(\sigma | x_i={\rm B}) \Pr(x_i={\rm B})} \nonumber \\
&= \dfrac{\sum_{s=a,b} \overline{\Pr}(\sigma | s) \Pr(s | x_i={\rm A}) \Pr(x_i={\rm A})}{\sum_{s=a,b} \overline{\Pr}(\sigma | s) \Pr(s | x_i={\rm A}) \Pr(x_i={\rm A}) + \sum_{s=a,b} \overline{\Pr}(\sigma | s) \Pr(s | x_i={\rm B}) \Pr(x_i={\rm B})} \nonumber \\
&=\begin{cases}
    \dfrac{\theta \Pr(x_i={\rm A})}{\theta \Pr(x_i={\rm A}) + (1-\theta) \Pr(x_i={\rm B})}     & (\sigma=\alpha), \\
    \dfrac{(1-\theta) \Pr(x_i={\rm A})}{(1-\theta) \Pr(x_i={\rm A}) + \theta \Pr(x_i={\rm B})} & (\sigma=\beta).  \\
\end{cases}
\label{eq:PiA_update}
\end{align} 
It should be noted that $\Pr(x_i={\rm B}|\sigma)=1-\Pr(x_i={\rm A}|\sigma)$. Then, we increment the time by $1/N$ such that each agent is updated once per unit time on average. Iterative application of Eq. (\ref{eq:PiA_update}) leads to   
\begin{align} 
\Pr(x_i={\rm A}) = \dfrac{\theta^{n_{\alpha i}} (1-\theta)^{n_{\beta i}}}{\theta^{n_{\alpha i}} (1-\theta)^{n_{\beta i}} + (1-\theta)^{n_{\alpha i}} \theta^{n_{\beta i}}} 
= \left\{ 1 + \left( \dfrac{1-\theta}{\theta} \right)^{n_{\alpha i} - n_{\beta i}} \right\}^{-1} \label{eq:PiA_specific_form}
\end{align}
and
\begin{align} 
\Pr(x_i={\rm B}) = \dfrac{(1-\theta)^{n_{\alpha i}} \theta^{n_{\beta i}}}{(1-\theta)^{n_{\alpha i}} \theta^{n_{\beta i}} + \theta^{n_{\alpha i}} (1-\theta)^{n_{\beta i}}}
= \left\{ 1 + \left( \dfrac{1-\theta}{\theta} \right)^{n_{\beta i} - n_{\alpha i}} \right\}^{-1}, 
\label{eq:PiB_specific_form}
\end{align}
where $n_{\alpha i}$ ($n_{\beta i}$) is the 
accumulated 
number of signals $\sigma=\alpha$ ($\sigma=\beta$) that agent $i$ has perceived. The state of each agent $i$ is uniquely determined by $n_{\alpha i}-n_{\beta i}$, which is consistent with basic Bayesian theory \cite{RabinSchrag1999QJECON, Perez-EscuderodePolavieja2011PLoSComputBiol}. 

\section{Results} \label{sec:Results}
\subsection{Setup for numerical simulations}
Unless otherwise stated, we set $1/2 < \theta < 1$ and assume a neutral initial condition $\Pr(x_i={\rm A})=1/2$ ($1\le i \le N$), or, equivalently, $n_{\alpha i} = n_{\beta i}$ ($1 \le i \le N$). The agents exchange signals and update their beliefs, possibly under a confirmation bias. After a transient, the agents believe in either opinion with a strong confidence, i.e., $\Pr(x_i={\rm A}) \approx 0$ or $1$. We halt a run when $|n_{\alpha i}-n_{\beta i}| \ge \Delta n_{\rm c}$ is satisfied for all $i$ for the first time, where $\Delta n_{\rm c}$ is the threshold. 
In other words, a run continues if at least one agent $i$ has the $|n_{\alpha i}-n_{\beta i}|$ value smaller than $\Delta n_{\rm c}$.

\subsection{Case without the confirmation bias}
We first consider the case without a confirmation bias (i.e., $q=0$). We investigate the dynamics of the mean belief $\overline{P}_{\rm A}(t) \equiv \sum_{i=1}^N \Pr(x_i={\rm A})/N$ at time $t$ by drawing a return map, i.e., $\overline{P}_{\rm A}(t)$ as a function of $\overline{P}_{\rm A}(t-1)$ \cite{Galam2002EurPhysJB,PachecoSantosSouzaSkyrms2009ProcRSocB,PachecoPinheiroSantos2009PLoSComputBiol}. The return map for $N=100$, $\theta=0.99$, and $\Delta n_{\rm c}=500$ based on $1000$ runs is shown in Fig. \ref{fig:meanP_ba}. Because $\overline{P}_{\rm A}(t) > \overline{P}_{\rm A}(t-1)$ when $0.5 < \overline{P}_{\rm A}(t-1) < 1$ and $\overline{P}_{\rm A}(t) < \overline{P}_{\rm A}(t-1)$ when $0 < \overline{P}_{\rm A}(t-1) < 0.5$, the dynamics is in accordance with majority rule behavior. All $1000$ runs finished with an agreement of opinion A [i.e., $\Pr(x_i={\rm A}) \approx 1$ for all $i$] or opinion B [i.e., $\Pr(x_i={\rm A})\approx 0$ for all $i$]. Each case occurred approximately half the time.

\subsection{Case with the confirmation bias}
We turn on the confirmation bias to examine the possibility that it induces disagreement among agents. At least for large $q$ (i.e., $q \approx 1$), disagreement is expected to be reached because the first perceived signal would determine the final belief of each agent and is equally likely to be $\alpha$ and $\beta$ for many agents. 

In the following numerical simulations, we measured the degree of disagreement, which we defined as follows. We determined that agreement was reached in a run if the final signs of $n_{\alpha i}-n_{\beta i}$ were the same for $i=1,\ldots,N$. Otherwise, we said that disagreement was reached. We denoted the fraction of runs that finished with disagreement by $F_{\rm d}$. 

\begin{figure}[t]
\includegraphics[width=1.0\linewidth]{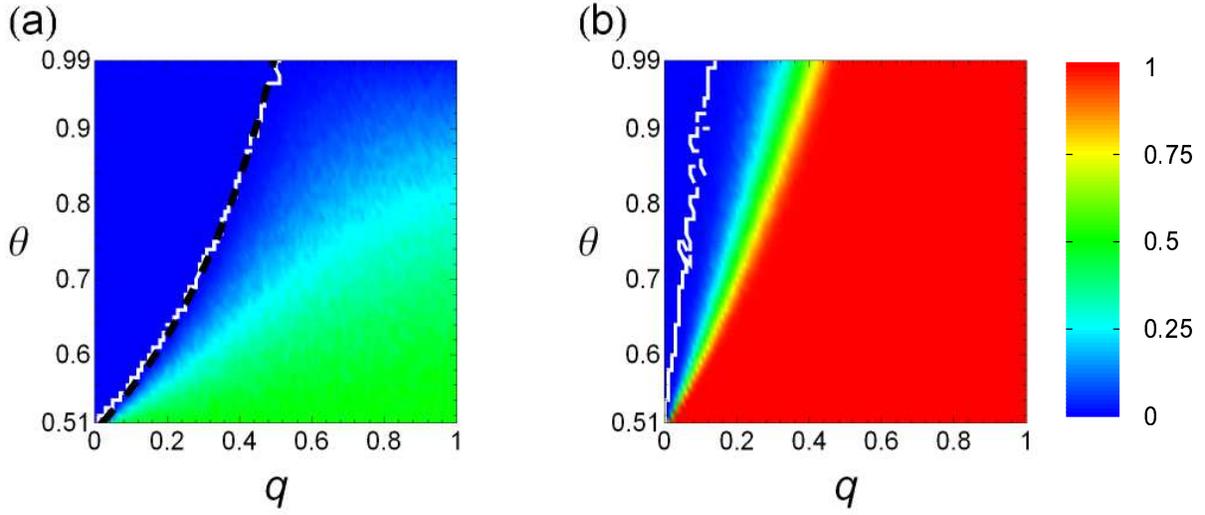}
\caption{
Fraction of disagreement $F_{\rm d}$. (a) $N=2$. (b) $N=100$. Solid lines represent the boundary between $F_{\rm d}=0$ and $F_{\rm d} > 0$. The dashed line 
in (a) 
represents $q=1-1/2\theta$. 
The dashed line is not drawn in (b) because this theoretical estimate is valid only for $N=2$. 
In (a), the two lines almost overlap each other. The initial belief of each agent was assumed to be neutral [i.e., $\Pr(x_i={\rm A}) = 0.5$, $1 \le i \le N$]. 
}
\label{fig:R_Bm}
\end{figure}

We set $\Delta n_{\rm c}=500$ and the number of runs to $1000$. In Figs. \ref{fig:R_Bm}(a) and \ref{fig:R_Bm}(b), $F_{\rm d}$ is shown as a function of $q$ and $\theta$ for $N=2$ and $100$, respectively.  
First, $F_{\rm d}$ monotonically increases with $q$ and decreases with $\theta$ for both $N=2$ and $N=100$. It should be noted that disagreement occurred in at least one run in the regions right to the solid fractured lines in Fig. \ref{fig:R_Bm}. Second, $F_{\rm d}$ for $N=2$ [Fig. \ref{fig:R_Bm}(a)] is smaller than $F_{\rm d}$ for $N=100$ [Fig. \ref{fig:R_Bm}(b)] for all the $q$ and $\theta$ values. Therefore, disagreement seems to be a likely outcome of the model for large $N$, particularly for large $q$ and small $\theta$. When $N=100$, perfect agreement, i.e., $F_{\rm d}=0$, is realized only for $q$ close to zero. In other words, even a small degree of confirmation bias elicits disagreement among the agents.

\subsection{Probability flow analysis for $N=2$}
To obtain analytical insights into the model, we performed an annealed approximation for $N=2$ by averaging out fluctuations of the dynamics for different times and runs. The configuration of the population is specified by $(m_1, m_2) \equiv (n_{\alpha 1}-n_{\beta 1}, n_{\alpha 2}-n_{\beta 2})$. The stochastic dynamics of the model can be mapped to a random walk on the two-dimensional lattice; a walker is initially located at $(m_1, m_2) = (0,0)$ and randomly hops to one of the four neighboring lattice points in each time step. We defined $f_R(m_1, m_2)$, $f_L(m_1, m_2)$, $f_U(m_1, m_2)$, and $f_D(m_1, m_2)$ as the probabilities that the walker located at $(m_1, m_2)$ moves to $(m_1+1, m_2)$, $(m_1-1, m_2)$, $(m_1, m_2+1)$, and $(m_1, m_2-1)$, respectively. 
The four probabilities are given by 
\begin{align} 
f_R(m_1, m_2)    
    =\begin{cases}
    \dfrac{(1-q) \Pr(s=a | m_2) + q}{2} &(m_1 \ge 1), \\[6pt]
    \dfrac{\Pr(s=a | m_2)}{2} &(m_1 = 0), \\[6pt]
    \dfrac{(1-q) \Pr(s=a | m_2)}{2} &(m_1 \le -1),
    \end{cases} 
\label{eq:fR}
\end{align}
\begin{align} 
f_L(m_1, m_2)=\dfrac{1}{2}-f_R(m_1, m_2),  
\label{eq:fL}
\end{align} 
\begin{align} 
f_U(m_1, m_2) 
    =\begin{cases}
    \dfrac{(1-q) \Pr(s=a | m_1) + q}{2} &(m_2 \ge 1), \\[6pt]
     \dfrac{\Pr(s=a | m_1)}{2} &(m_2 = 0), \\[6pt]
        \dfrac{(1-q) \Pr(s=a | m_1)}{2} &(m_2 \le -1),
    \end{cases} 
\label{eq:fU}
\end{align}
and 
\begin{align} 
f_D(m_1, m_2)=\dfrac{1}{2}-f_U(m_1, m_2),  
\label{eq:fD}
\end{align}
where 
\begin{align} 
\Pr(s=a|m_j)  
&= 1 - \theta + (2 \theta - 1) \Pr(x_j={\rm A}|m_j) \nonumber \\
&= 1 - \theta + (2 \theta - 1) \left\{ 1 + \left( \dfrac{1-\theta}{\theta} \right)^{m_j} \right\}^{-1}
\label{eq:Pja_rj}
\end{align} 
is the probability that agent $j$ with $n_{\alpha j}-n_{\beta j}=m_j$ imparts signal $s=a$.
$f_R(m_1, m_2)$ and $f_U(m_1, m_2)$ increase with $m_1$ and $m_2$, and $f_L(m_1, m_2)$ and $f_D(m_1, m_2)$ decrease with $m_1$ and $m_2$. 

In the following, we study the mean dynamics of the random walk driven by the drift terms. Because the transition probability of the random walk is symmetric with respect to the lines $m_1 = m_2$ and $m_1 = -m_2$, we focus on the region given by $-m_2 \le m_1 \le m_2, m_2 > 0$. 
We define $m_{1 \rm c}$ and $m_{2 \rm c}$, which are not integers in general, as the values satisfying $f_U(m_{1 \rm c}, m_2) = f_D(m_{1 \rm c}, m_2) \ ^\forall m_2 >0$ and $f_R(m_1, m_{2 \rm c}) = f_L(m_1, m_{2 \rm c}) \ ^\forall m_1 < 0$, respectively. They are given by
\begin{align} 
m_{2 \rm c} &= - m_{1 \rm c} \nonumber \\
&= \dfrac{\ln \left[ (2 \theta - 1) \left\{ \dfrac{1}{2(1-q)} - (1-\theta) \right\}^{-1} -1 \right]^{-1}}{\ln \dfrac{\theta}{1-\theta}}.  
\label{eq:r2_fR=fL}
\end{align}
Note that $m_{1 \rm c}$ and $m_{2 \rm c}$ exist if and only if $(2 \theta - 1) \left\{ 1/2(1-q) - (1-\theta) \right\}^{-1} -1 > 0$, i.e., 
\begin{align} 
q < 1 - \dfrac{1}{2 \theta}.  
\label{eq:q_1-1_2theta}
\end{align}

\begin{figure}[t]
\includegraphics[width=1.0\linewidth]{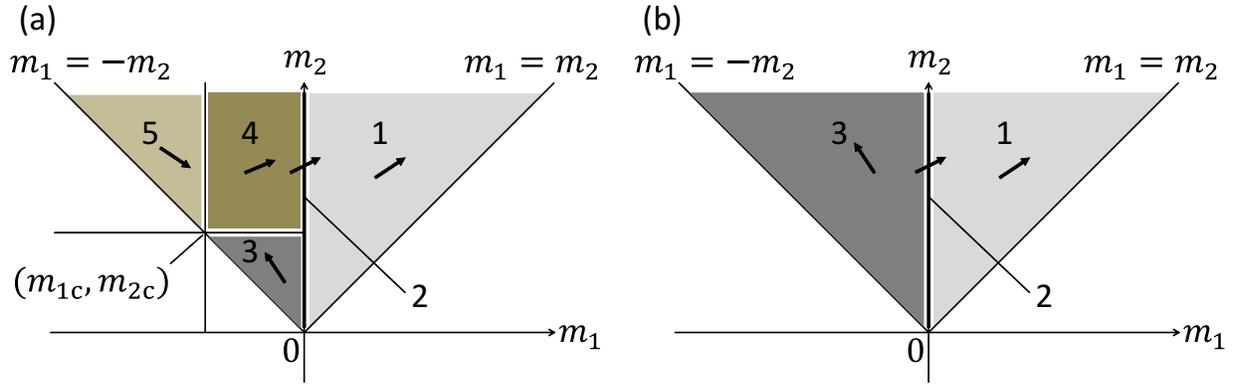}
\caption{
Schematic of the probability flow given  by $f_R(m_1,m_2)-f_L(m_1,m_2)$ and $f_U(m_1,m_2)-f_D(m_1,m_2)$. (a) $q < 1-1/2\theta$. (b) $q > 1-1/2\theta$. The labels from $1$ to $5$ correspond to the five regions.
}
\label{fig:schematicview_fR-fL_fU-fD}
\end{figure}

\begin{figure}[t]
\includegraphics[width=1.0\linewidth]{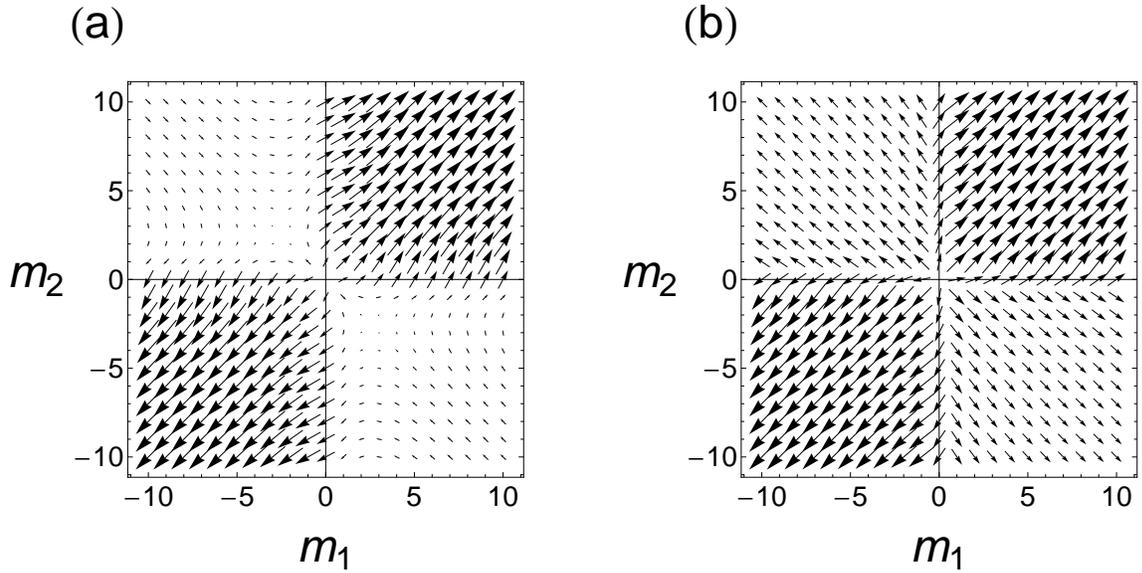}
\caption{
Probability flow of the opinion dynamics when $N=2$. Vector $(f_R(m_1,m_2)-f_L(m_1,m_2),f_U(m_1,m_2)-f_D(m_1,m_2))$ is shown by an arrow of proportional size at each position of the random walker $(m_1,m_2)$. (a) $q=0.15$ and $\theta=0.64$, which satisfies $q<1-1/2\theta$. (b) $q=0.4$ and $\theta=0.64$, which satisfies $q>1-1/2\theta$. The size of the vectors is manually normalized for clarity, independently for the two panels.
}
\label{fig:vectorplot}
\end{figure}

First, we consider the case $q < 1 - 1/2 \theta$. We partition the upper quadrant of the lattice (given by $-m_2 \le m_1 \le m_2, m_2 > 0$) into five regions: 
region $1$ ($0 < m_1 \le m_2$), 
region $2$ ($m_1=0, m_2 > 0$), 
region $3$ ($-m_2 \le m_1 < 0, 0 < m_2 < m_{2 \rm c}$), 
region $4$ ($m_{1 \rm c} < m_1 < 0, \ m_2 > m_{2 \rm c}$), and
region $5$ ($-m_2 \le m_1 < m_{1 \rm c}$), 
as shown in Fig. \ref{fig:schematicview_fR-fL_fU-fD}(a).
We obtain from the condition $1/2 < \theta <1$  
\begin{align} 
f_R(m_1, m_2) - f_L(m_1, m_2) \ge f_U(m_1, m_2) - f_D(m_1, m_2) > 0 
\label{eq:fRfLfUfD}
\end{align}
in region $1$, 
\begin{align} 
f_R(m_1, m_2) - f_L(m_1, m_2) > 0, \ \ f_U(m_1, m_2) - f_D(m_1, m_2) = \dfrac{q}{2} 
\label{eq:fRfLfUfD2}
\end{align}
in region $2$, 
\begin{align} 
f_U(m_1, m_2) - f_D(m_1, m_2) \ge f_L(m_1, m_2) - f_R(m_1, m_2) > 0
\label{eq:fRfLfUfD3}
\end{align}
in region $3$, 
\begin{align} 
f_R(m_1, m_2) - f_L(m_1, m_2) > 0, \ \ f_U(m_1, m_2) - f_D(m_1, m_2) > 0
\label{eq:fRfLfUfD4}
\end{align}
in region $4$, and
\begin{align} 
f_R(m_1, m_2) - f_L(m_1, m_2) \ge f_D(m_1, m_2) - f_U(m_1, m_2) > 0
\label{eq:fRfLfUfD5}
\end{align}
in region $5$. 
The probability flow of the walker after the annealed approximation, i.e., ($f_R(m_1, m_2) - f_L(m_1, m_2), f_U(m_1, m_2) - f_D(m_1, m_2)$) inferred from Eqs. \eqref{eq:fRfLfUfD}-\eqref{eq:fRfLfUfD5} is shown schematically in Fig. \ref{fig:schematicview_fR-fL_fU-fD}(a). If the walker is in the second quadrant (i.e., regions $3$, $4$, and $5$) where the two agents disagree with each other, the random walker is likely to eventually escape and enter the first quadrant (i.e., region $1$) where the two agents agree with each other. In fact, Fig. \ref{fig:vectorplot}(a), which shows the actual probability flow, indicates that the agreement necessarily occurs. Therefore, agreement is the expected outcome when $q<1-1/2\theta$. 

Second, if $q > 1 - 1/2 \theta$, regions $4$ and $5$ are absent because $m_{1c}$ and $m_{2c}$ diverge. Regions $1$ and $2$, in which inequalities \eqref{eq:fRfLfUfD} and \eqref{eq:fRfLfUfD2} are satisfied, respectively, are the same as those in the case $q<1-1/2\theta$. Region $3$, in which inequality \eqref{eq:fRfLfUfD3} is satisfied, is modified to $-m_2 \le m_1 < 0$. The probability flows are schematically shown in Fig. \ref{fig:schematicview_fR-fL_fU-fD}(b). 
$f_L(m_1, m_2)>f_R(m_1, m_2)$ and $f_U(m_1, m_2)>f_D(m_1, m_2)$ are satisfied in region $3$. Therefore, once the walker is deep in the second quadrant, it is likely to move toward $m_1 \to -\infty$ and $m_2 \to \infty$, which implies that two agents finally disagree. The actual probability flow shown in Fig. \ref{fig:vectorplot}(b) is consistent with this prediction.

The transition line $q = 1 - 1/2 \theta$ is shown by the dashed line in Fig. \ref{fig:R_Bm}(a). It accurately predicts the parameter region in which disagreement can occur, i.e., the region right to the solid line.

The same transition line is also derived for the Rabin-Schrag model, which is concerned with a single agent subjected to a confirmation bias \cite{RabinSchrag1999QJECON}. In their model, the agent forms a  belief by repetitively receiving a stochastic signal $s \in \left\{ a,b \right\}$ from nature, according to $\Pr(s=a|x={\rm A})=(s=b|x={\rm B})=\theta$. Rabin and Schrag calculated the probability that the agent eventually misunderstands the state of the nature (i.e., A or B), starting from neutral belief. This probability is equal to zero when $q \le 1-1/2\theta$ and positive when $q > 1-1/2\theta$ (see proposition $4$ in \cite{RabinSchrag1999QJECON}). Our results obtained in this section are consistent with theirs because disagreement in our model roughly corresponds to misunderstanding in the Rabin-Schrag model. 

\subsection{Different disagreement configurations for large $N$}
In general, there are $N-1$ disagreement configurations, as distinguished by the number of agents that finally believe in opinion A, which ranges from $1$ to $N-1$. To distinguish different disagreement configurations, we examined the fraction of agents that believed in the minority opinion at the end of a run. We averaged this fraction over the runs ending with disagreement. We called this quantity the average size of the minority. 

\begin{figure}[t]
\includegraphics[width=1.0\linewidth]{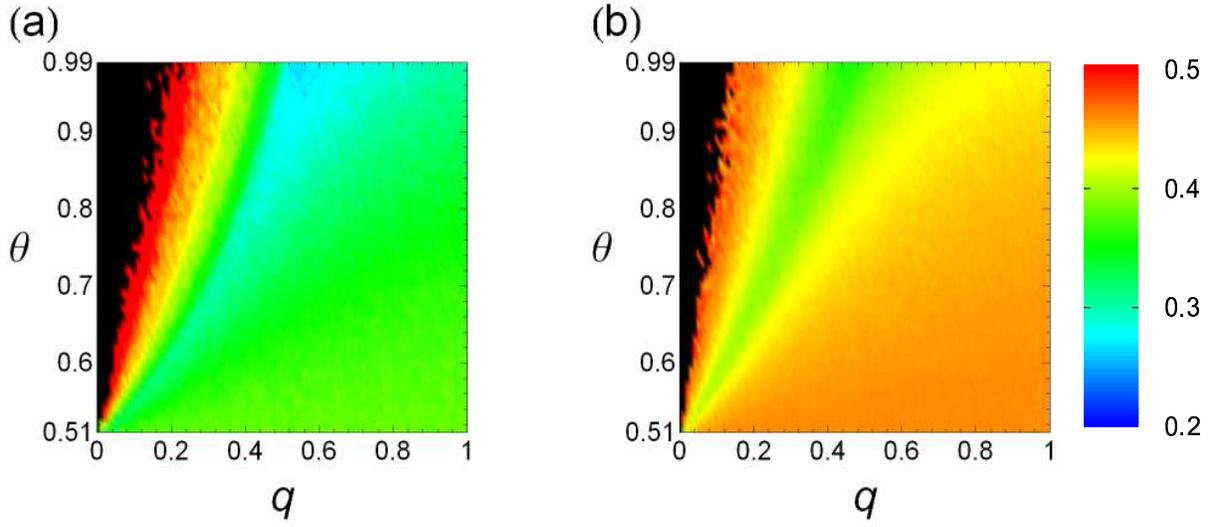}
\caption{
Average size of the minority. (a) $N=10$ and (b) $N=100$. The initial belief of each agent is assumed to be neutral [i.e., $\Pr(x_i={\rm A}) = 0.5$, $1 \le i \le N$]. The black region represents the case where all the $1000$ runs end with agreement such that the average size of the minority is undefined.
}
\label{fig:R_Minority_MinorityMin}
\end{figure}

Figures \ref{fig:R_Minority_MinorityMin}(a) and \ref{fig:R_Minority_MinorityMin}(b) show the average size of the minority for $N=10$ and $N=100$, respectively. The black regions indicate the parameter values for which the average size of the minority is undefined because all $1000$ runs end with agreement.  
When $q$ is small, the average size of the minority monotonically decreases with $q$ and monotonically increases with $\theta$ for both $N=10$ and $100$. Therefore, small $q$ and large $\theta$ values allow only balanced disagreement configurations, in which the numbers of the agents believing in the opposite opinions are close to $N/2$. 

However, the average size of the minority increases when $q$ is large. This is particularly the case for $N=100$ [Fig. \ref{fig:R_Minority_MinorityMin}(b)]. This increase occurs for the following reason. With a strong confirmation bias, agents end up with an opinion consistent with a small number of signals perceived in the early stages, and both signals are equally likely to be observed in the early stages under neutral initial conditions. In the extreme case in which $q=1$, agents reinforce the opinion that is consistent with their first perceived signal. Therefore, unbalanced disagreement configurations are rarely realized when $q$ is large. 

\subsection{Effects of the system size and initial condition}
Figures \ref{fig:R_Bm} and \ref{fig:R_Minority_MinorityMin} suggest that the agreement is unlikely to be reached in a large population. To examine the effect of the population size, we defined $q_{\rm c}$ as the value of $q$ such that a threshold number of runs among $10^4$ runs end with agreement. For a given $\theta$ value, we determined $q_{\rm c}$ by the bisection method. The number of agreement runs may not monotonically change in $q$ because the number of runs is finite. Therefore, the bisection method does not perfectly work in general. However, we corroborated that the following results were negligibly affected by the lack of monotonicity. 
 
\begin{figure}[t]
\begin{tabular}{cc}
\begin{minipage}[t]{.5\hsize}
\begin{center}
\includegraphics[width=\hsize]{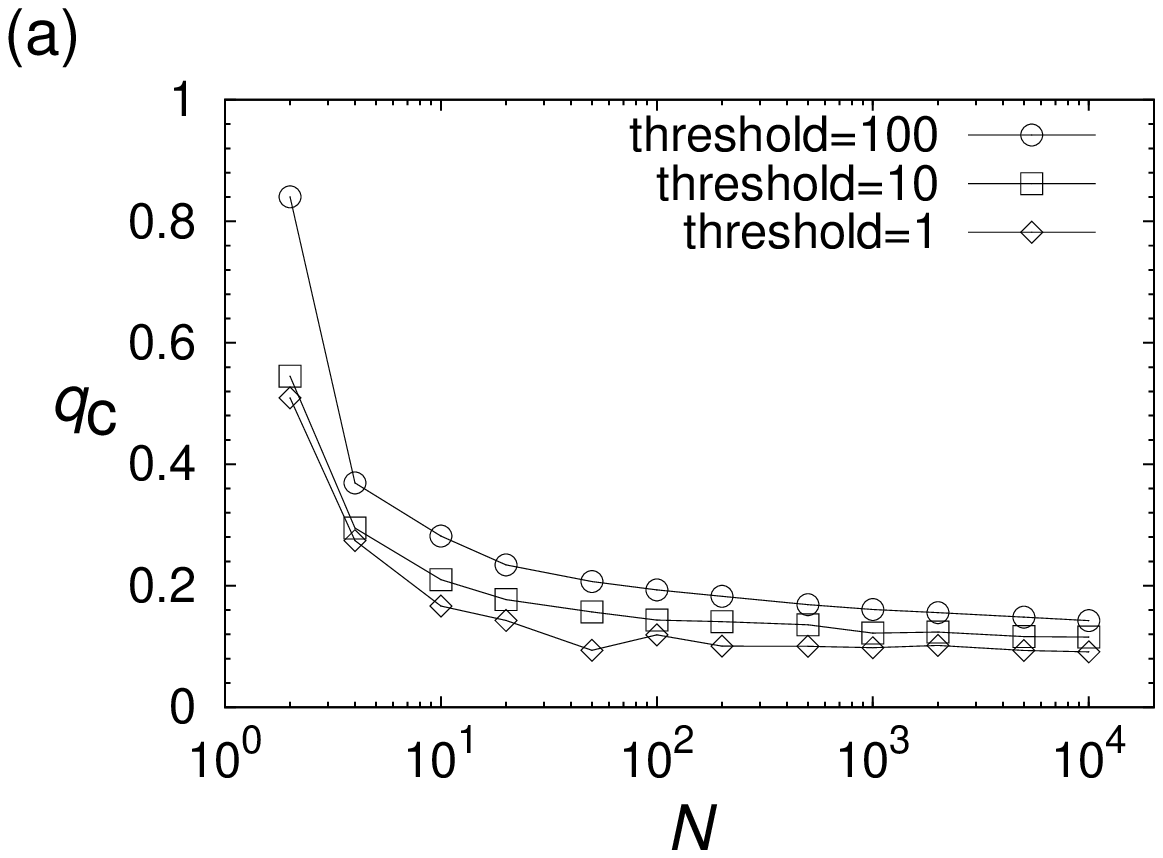}
\end{center}
\end{minipage}
\begin{minipage}[t]{.5\hsize}
\begin{center}
\includegraphics[width=\hsize]{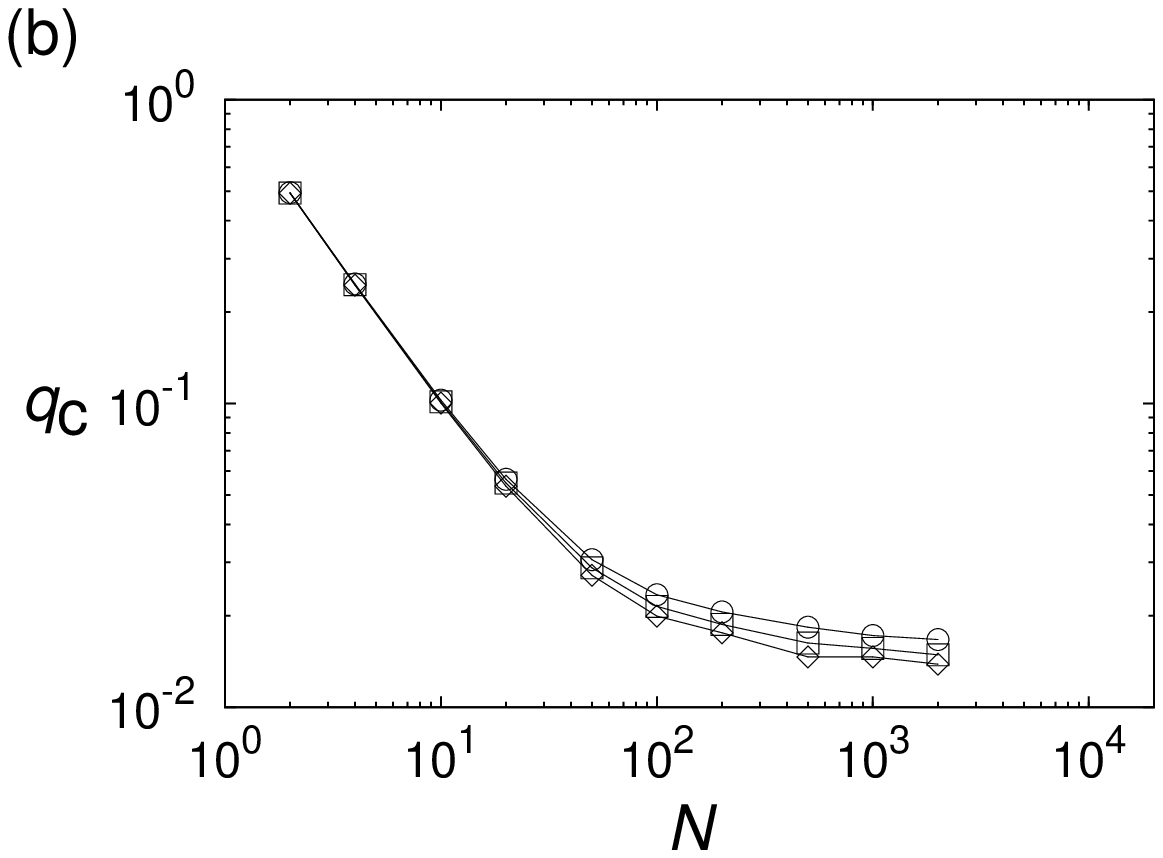}
\end{center}
\end{minipage}
\end{tabular}
\begin{tabular}{cc}
\begin{minipage}[t]{.5\hsize}
\begin{center}
\includegraphics[width=\hsize]{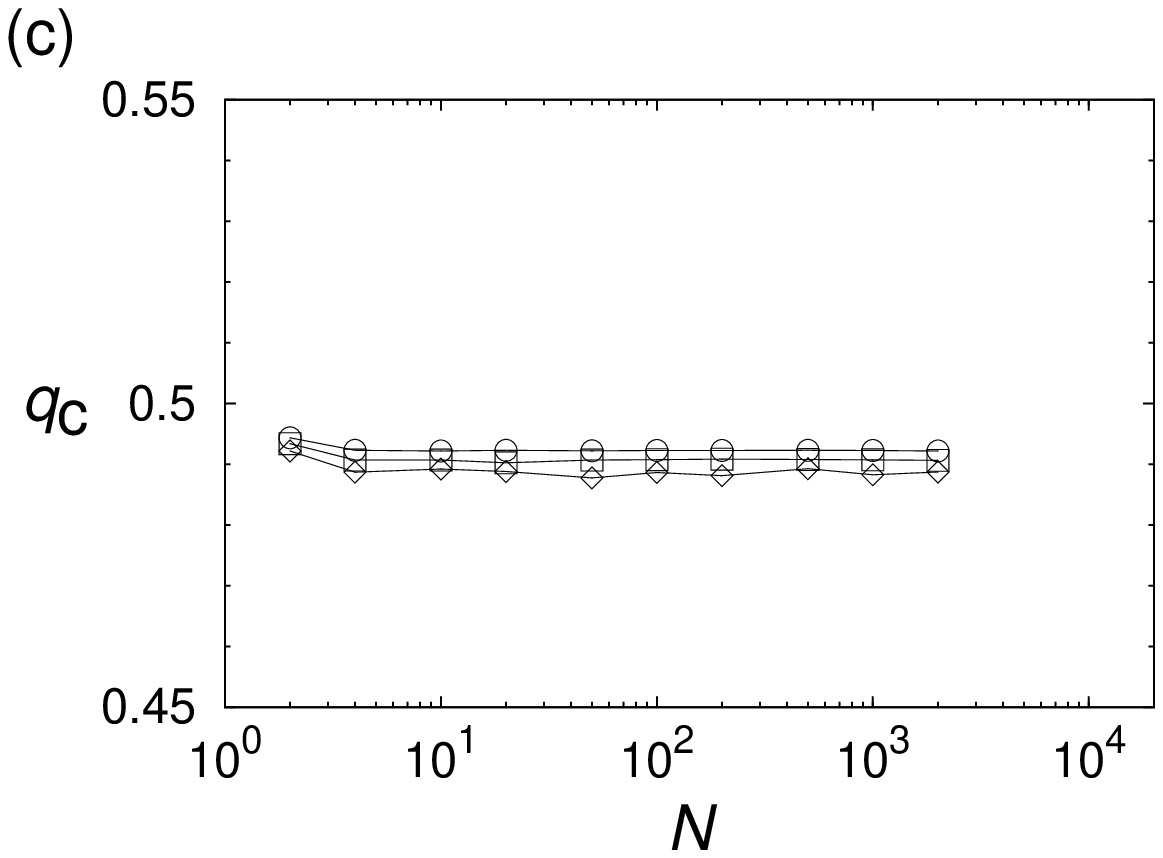}
\end{center}
\end{minipage}
\begin{minipage}[t]{.5\hsize}
\begin{center}
\end{center}
\end{minipage}
\end{tabular}
\caption{
Threshold for the agreement-disagreement transition ($q_{\rm c}$) as a function of $N$ under the (a) neutral, (b) bimodal, and (c) most unbalanced initial conditions. We set $\theta=0.99$.
}
\label{fig:qcr_vs_N}
\end{figure}

The dependence of $q_{\rm c}$ on $N$ is shown in Fig. \ref{fig:qcr_vs_N}(a) for three threshold values. For example, the results for the threshold value $100$ (shown by circles) indicate that at least $100$ runs among the $10^4$ runs end up with disagreement when 
$q > q_{\rm c}$. 
We set $\theta=0.99$ and $\Delta n_{\rm c}=500$. To explore the possibility of disagreement in large populations, we set $\theta$ close to $1$. It should be noted that Fig. \ref{fig:R_Bm} indicates that the probability of disagreement is small for a large $\theta$ value. In Fig. \ref{fig:qcr_vs_N}(a), $q_{\rm c}$ quickly decreases for $N \le 10$ and gradually decreases for $N \ge 100$. Disagreement often occurs for large $N$ unless $q$ is small. Nevertheless, Fig. \ref{fig:qcr_vs_N}(a) suggests that the range of $q$ for which agreement always occurs survives for diverging $N$.

In generating Fig. \ref{fig:qcr_vs_N}(a), we used an initial condition in which all the agents had a neutral belief [i.e., $\Pr(x_i={\rm A})=0.5, 1 \le i \le N$]. To check the effect of the initial condition, we investigated the dependence of $q_{\rm c}$ on $N$ under two other initial conditions. In the bimodal initial condition, we initially set $n_{\alpha i} - n_{\beta i} = 100$ ($1 \le i \le N/2$) and $n_{\alpha i} - n_{\beta i} = -100$ ($N/2+1 \le i \le N$). We assumed that $N$ was even for this initial condition. In the so-called most unbalanced initial condition, we set $n_{\alpha 1} - n_{\beta 1} = 100$ and $n_{\alpha i} - n_{\beta i} = -100$ ($2 \le i \le N$).

The numerical results for the two initial conditions are shown in Figs. \ref{fig:qcr_vs_N}(b) and \ref{fig:qcr_vs_N}(c). The parameter values $\theta=0.99$, $\Delta n_{\rm c}=500$ are the same as those used in Fig. \ref{fig:qcr_vs_N}(a). The transition point $q_{\rm c}$ decreases with $N$ more rapidly with the bimodal initial condition [Fig. \ref{fig:qcr_vs_N}(b)] than with the neutral initial condition [Fig. \ref{fig:qcr_vs_N}(a)]. This result is intuitive: the bimodal initial condition paves the way to disagreement. In contrast, $q_{\rm c}$ under the most unbalanced initial condition is almost constant near $0.5$ irrespective of $N$. Therefore, disagreement is highly unlikely unless the confirmation bias is strong (i.e., $q$ is greater than $0.5$). The results shown in Fig. \ref{fig:qcr_vs_N} suggest that the eventual behavior of the model strongly depends on the initial condition even after the results are averaged over runs. 

\section{Discussion} \label{sec:Discussion}
\begin{figure}[t]
\includegraphics[width=0.6\linewidth]{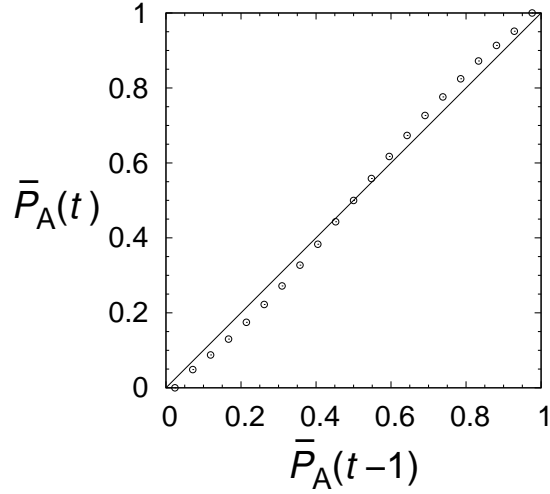}
\caption{
Return map of the mean belief. We set $N=100$, $q=0$, and $\theta=0.99$. We recorded the values of $(\overline{P}_{\rm A}(t-1), \overline{P}_{\rm A}(t)$) for $t = 1,1+1/N,1+2/N,\ldots$ for $1000$ runs and divided the recorded pairs into $21$ classes. The $k$th class ($k=1,\ldots,21$) was composed of the pairs satisfying $(k-1)/21 \le \overline{P}_{\rm A}(t-1) < k/21$. We obtained the mean value $\left< \overline{P}_{\rm A}(t) \right>_k$ for the $k$th class by averaging $\overline{P}_{\rm A}(t)$ over all the pairs contained in the $k$th class. Finally, we plotted $\left< \overline{P}_{\rm A}(t) \right>_k$ against $(k-0.5)/21$ for $k=1,\ldots,21$. The diagonal $\overline{P}_{\rm A}(t)=\overline{P}_{\rm A}(t-1)$ is also shown as a guide.
}
\label{fig:meanP_ba}
\end{figure}

Our numerical results are summarized as follows. When the confirmation bias is absent (i.e., $q=0$), the opinion dynamics under the Bayesian update rule leads to the complete agreement among agents. The behavior of the model is similar to majority rule dynamics (Fig. \ref{fig:meanP_ba}). 
When the confirmation bias is present, disagreement is a likely outcome, particularly for a strong confirmation bias (i.e., large $q$). Disagreement is also more likely for a lower fidelity of the signal (i.e., $\theta \approx 1/2$) and a larger system size. The transition line separating the parameter region in which both agreement and disagreement can occur and that in which only agreement occurs is approximately given by $q=1-1/2\theta$ when $N=2$. This line is identical to the one determined by Rabin and Schrag for their model for a single agent's decision making \cite{RabinSchrag1999QJECON}. Finally, the behavior of the model strongly depends on the initial condition.

Our model and results are different from Orl\'{e}an's \cite{Orlean1995JEconBehavOrgan}, although Orl\'{e}an's model employs multiple agents that perform the Bayesian updates under a confirmation bias. First, the belief of each agent is binary in Orl\'{e}an's model, whereas our model introduces an infinite range of discrete beliefs, as in \cite{RabinSchrag1999QJECON}. Second, interaction between agents is introduced differently in the two models. In Orl\'{e}an's model, each agent refers to the global fraction of agents believing in one of the two opinions. In our model, agents refer to other opinions by peer-to-peer interaction, i.e., by receiving a binary signal that is correlated with the belief of the sender. Third, the stochastic dynamics of Orl\'{e}an's model is ergotic when the collective opinion does not reach agreement. The collective opinion obeys a stationary distribution, irrespective of the initial condition. In contrast, in all our simulations, the stochastic dynamics of our model was nonergotic, such that the final configuration depended on the initial condition in a wide parameter region.   

In social science studies of polarization, several authors analyzed Bayesian models in which different agents receiving a series of common signals end up in disagreement. The proposed mechanisms governing disagreement include different initial beliefs or factors that affect perception of later incoming signals \cite{AndreoniMylovanov2012AmEconJ-Microecon,DixitWeibull2007PNAS,ZimperLudwig2009JRiskUncertain,AcemogluVictorMuhamet2009Fragility,AcemogluOzdaglar2011DynGamesAppl}, different update rules \cite{ZimperLudwig2009JRiskUncertain}, and ambiguity aversion \cite{ZimperLudwig2009JRiskUncertain,BaligaHananyKlibanoff2011Polarization}. These models and ours are different in three major ways. First, a ground truth opinion corresponding to the state of nature is assumed in these models but not in ours. Second, public signals commonly received by different agents are assumed in these models but not in ours. Third, the agents do not have direct peer-to-peer interaction in these models, but they do in ours. Models with interacting Bayesian agents, which show disagreement (reviewed in Ref. \cite{AcemogluOzdaglar2011DynGamesAppl}), are also different from our model in the first respect. It should be noted that Zimper and Ludwig discussed confirmation bias with their Bayesian model \cite{ZimperLudwig2009JRiskUncertain}. However, they derived a confirmation bias from their model, rather than assuming one, such that their results pertaining to confirmation bias were also distinct from ours. 

Extending our model to the case of networks is straightforward. For example, we can select a recipient of the signal with probability $1/N$ and then select the sender with equal probability among the neighbors of the recipient on the network. Another possible update rule is to select the sender first and then the recipient among the sender's neighbors. Yet another possibility is to select a link with equal probability and designate one of the two agents as sender and the other as recipient. On heterogeneous networks, the results may depend on the update rule because it is the case in the voter model \cite{AntalRednerSood2006PhysRevLett,SoodAntalRedner2008PhysRevE,MasudaOhtsuki2009NewJPhys,Masuda2009JTheorBiol}. Extension of the model to the case of confirmation bias heterogeneity may also be interesting. Neurological evidence shows that different individuals have different confirmation bias strengths \cite{DollHutchisonFrank2011JNeurosci}. The strength of the confirmation bias and the position of the node in a social network may be correlated and affect the dynamics. It is also straightforward to extend the model to the case of multiple opinion cases. These and other extensions, along with the study of analytically tractable models that capture the essence of the present study, warrant future work.

\begin{acknowledgments}
We thank Mitsuhiro Nakamura, Taro Takaguchi, and Shoma Tanabe for critical reading of the manuscript. This work is supported by Grants-in-Aid for Scientific Research [Grant No. 23681033 and Innovative Areas ``Systems Molecular Ethology" (Grant No. 20115009)] from MEXT, Japan.
\end{acknowledgments}

\clearpage

\clearpage

\clearpage

\clearpage

\clearpage

\clearpage

\clearpage

\clearpage


\begin{thebibliography}{99}%
\makeatletter
\providecommand \@ifxundefined [1]{%
 \@ifx{#1\undefined}
}%
\providecommand \@ifnum [1]{%
 \ifnum #1\expandafter \@firstoftwo
 \else \expandafter \@secondoftwo
 \fi
}%
\providecommand \@ifx [1]{%
 \ifx #1\expandafter \@firstoftwo
 \else \expandafter \@secondoftwo
 \fi
}%
\providecommand \natexlab [1]{#1}%
\providecommand \enquote  [1]{``#1''}%
\providecommand \bibnamefont  [1]{#1}%
\providecommand \bibfnamefont [1]{#1}%
\providecommand \citenamefont [1]{#1}%
\providecommand \href@noop [0]{\@secondoftwo}%
\providecommand \href [0]{\begingroup \@sanitize@url \@href}%
\providecommand \@href[1]{\@@startlink{#1}\@@href}%
\providecommand \@@href[1]{\endgroup#1\@@endlink}%
\providecommand \@sanitize@url [0]{\catcode `\\12\catcode `\$12\catcode
  `\&12\catcode `\#12\catcode `\^12\catcode `\_12\catcode `\%12\relax}%
\providecommand \@@startlink[1]{}%
\providecommand \@@endlink[0]{}%
\providecommand \url  [0]{\begingroup\@sanitize@url \@url }%
\providecommand \@url [1]{\endgroup\@href {#1}{\urlprefix }}%
\providecommand \urlprefix  [0]{URL }%
\providecommand \Eprint [0]{\href }%
\providecommand \doibase [0]{http://dx.doi.org/}%
\providecommand \selectlanguage [0]{\@gobble}%
\providecommand \bibinfo  [0]{\@secondoftwo}%
\providecommand \bibfield  [0]{\@secondoftwo}%
\providecommand \translation [1]{[#1]}%
\providecommand \BibitemOpen [0]{}%
\providecommand \bibitemStop [0]{}%
\providecommand \bibitemNoStop [0]{.\EOS\space}%
\providecommand \EOS [0]{\spacefactor3000\relax}%
\providecommand \BibitemShut  [1]{\csname bibitem#1\endcsname}%
\let\auto@bib@innerbib\@empty
\bibitem [{\citenamefont {Castellano}\ \emph {et~al.}(2009)\citenamefont
  {Castellano}, \citenamefont {Fortunato},\ and\ \citenamefont
  {Loreto}}]{CastellanoFortunatoLoreto2009RevModPhys}%
  \BibitemOpen
  \bibfield  {author} {\bibinfo {author} {\bibfnamefont {C.}~\bibnamefont
  {Castellano}}, \bibinfo {author} {\bibfnamefont {S.}~\bibnamefont
  {Fortunato}}, and\ \bibinfo {author} {\bibfnamefont {V.}~\bibnamefont
  {Loreto}},\ }\href@noop {} {\bibfield  {journal} {\bibinfo  {journal} {Rev.
  Mod. Phys.}\ }\textbf {\bibinfo {volume} {81}},\ \bibinfo {pages} {591}
  (\bibinfo {year} {2009})}\BibitemShut {NoStop}%
\bibitem [{\citenamefont {Krapivsky}\ \emph {et~al.}(2010)\citenamefont
  {Krapivsky}, \citenamefont {Redner},and\ \citenamefont
  {Ben-Naim}}]{KrapivskyRednerBen-Naim2010AKineticView}%
  \BibitemOpen
  \bibfield  {author} {\bibinfo {author} {\bibfnamefont {P.~L.}\ \bibnamefont
  {Krapivsky}}, \bibinfo {author} {\bibfnamefont {S.}~\bibnamefont {Redner}}, and\ 
  \bibinfo {author} {\bibfnamefont {E.}~\bibnamefont {Ben-Naim}},\
  }\href@noop {} 
  {\bibinfo {title} {\textit{A Kinetic View of Statistical Physics}}}\ 
  (\bibinfo  {publisher} {Cambridge University Press},\ \bibinfo
  {address} {Cambridge},\ \bibinfo {year} {2010})\BibitemShut {NoStop}%
\bibitem [{\citenamefont {Kuran}(1995)}]{Kuran1995PrivateTruthsPublicLies}%
  \BibitemOpen
  \bibfield  {author} {\bibinfo {author} {\bibfnamefont {T.}~\bibnamefont
  {Kuran}},\ }\href@noop {} 
  {\bibinfo {title} {\textit{Private Truths, Public Lies: The Social Consequences of Preference Falsification}}}\ 
  (\bibinfo  {publisher} {Harvard University Press},\ \bibinfo
  {address} {Cambridge, MA},\ \bibinfo {year} {1995})\BibitemShut {NoStop}%
\bibitem [{\citenamefont {Huckfeldt}\ \emph {et~al.}(2004)\citenamefont
  {Huckfeldt}, \citenamefont {Johnson},and\ \citenamefont
  {Sprague}}]{HuckfeldtJohnsonSprague2004PoliticalDisagreement}%
  \BibitemOpen
  \bibfield  {author} {\bibinfo {author} {\bibfnamefont {R.}~\bibnamefont
  {Huckfeldt}}, \bibinfo {author} {\bibfnamefont {P.~E.}\ \bibnamefont
  {Johnson}}, and\ \bibinfo {author} {\bibfnamefont {J.}~\bibnamefont
  {Sprague}},\ }\href@noop {} 
  {\bibinfo {title} {\textit{Political Disagreement: The Survival of Diverse Opinions within Communication Networks}}}\
  (\bibinfo  {publisher} {Cambridge University Press},\ \bibinfo
  {address} {Cambridge},\ \bibinfo {year} {2004})\BibitemShut {NoStop}%
\bibitem [{\citenamefont {Donnelly} and\ \citenamefont
  {Welsh}(1983)}]{DonnellyWelsh1983MathProcCambPhilSoc}%
  \BibitemOpen
  \bibfield  {author} {\bibinfo {author} {\bibfnamefont {P.}~\bibnamefont
  {Donnelly}} and\ \bibinfo {author} {\bibfnamefont {D.}~\bibnamefont
  {Welsh}},\ }\href@noop {} {\bibfield  {journal} {\bibinfo  {journal} {Math.
  Proc. Cambridge Philos. Soc.}\ }\textbf {\bibinfo {volume} {94}},\ \bibinfo {pages}
  {167} (\bibinfo {year} {1983})}\BibitemShut {NoStop}%
\bibitem [{\citenamefont {Sood} and\ \citenamefont
  {Redner}(2005)}]{SoodRedner2005PhysRevLett}%
  \BibitemOpen
  \bibfield  {author} {\bibinfo {author} {\bibfnamefont {V.}~\bibnamefont
  {Sood}} and\ \bibinfo {author} {\bibfnamefont {S.}~\bibnamefont {Redner}},\
  }\href@noop {} {\bibfield  {journal} {\bibinfo  {journal} {Phys. Rev. Lett.}\
  }\textbf {\bibinfo {volume} {94}},\ \bibinfo {pages} {178701} (\bibinfo
  {year} {2005})}\BibitemShut {NoStop}%
\bibitem [{\citenamefont {Antal}\ \emph {et~al.}(2006)\citenamefont {Antal},
  \citenamefont {Redner},and\ \citenamefont
  {Sood}}]{AntalRednerSood2006PhysRevLett}%
  \BibitemOpen
  \bibfield  {author} {\bibinfo {author} {\bibfnamefont {T.}~\bibnamefont
  {Antal}}, \bibinfo {author} {\bibfnamefont {S.}~\bibnamefont {Redner}}, and\ 
  \bibinfo {author} {\bibfnamefont {V.}~\bibnamefont {Sood}},\ }\href@noop
  {} {\bibfield  {journal} {\bibinfo  {journal} {Phys. Rev. Lett.}\ }\textbf
  {\bibinfo {volume} {96}},\ \bibinfo {pages} {188104} (\bibinfo {year}
  {2006})}\BibitemShut {NoStop}%
\bibitem [{\citenamefont {Sood}\ \emph {et~al.}(2008)\citenamefont {Sood},
  \citenamefont {Antal},and\ \citenamefont
  {Redner}}]{SoodAntalRedner2008PhysRevE}%
  \BibitemOpen
  \bibfield  {author} {\bibinfo {author} {\bibfnamefont {V.}~\bibnamefont
  {Sood}}, \bibinfo {author} {\bibfnamefont {T.}~\bibnamefont {Antal}}, and\
  \bibinfo {author} {\bibfnamefont {S.}~\bibnamefont {Redner}},\ }\href@noop {}
  {\bibfield  {journal} {\bibinfo  {journal} {Phys. Rev. E}\ }\textbf {\bibinfo
  {volume} {77}},\ \bibinfo {pages} {041121} (\bibinfo {year}
  {2008})}\BibitemShut {NoStop}%
\bibitem [{\citenamefont {Masuda} and\ \citenamefont
  {Ohtsuki}(2009)}]{MasudaOhtsuki2009NewJPhys}%
  \BibitemOpen
  \bibfield  {author} {\bibinfo {author} {\bibfnamefont {N.}~\bibnamefont
  {Masuda}} and\ \bibinfo {author} {\bibfnamefont {H.}~\bibnamefont
  {Ohtsuki}},\ }\href@noop {} {\bibfield  {journal} {\bibinfo  {journal} {New
  J. Phys.}\ }\textbf {\bibinfo {volume} {11}},\ \bibinfo {pages} {033012}
  (\bibinfo {year} {2009})}\BibitemShut {NoStop}%
\bibitem [{\citenamefont {Galam}(2002)}]{Galam2002EurPhysJB}%
  \BibitemOpen
  \bibfield  {author} {\bibinfo {author} {\bibfnamefont {S.}~\bibnamefont
  {Galam}},\ }\href@noop {} {\bibfield  {journal} {\bibinfo  {journal} {Eur.
  Phys. J. B}\ }\textbf {\bibinfo {volume} {25}},\ \bibinfo {pages} {403}
  (\bibinfo {year} {2002})}\BibitemShut {NoStop}%
\bibitem [{\citenamefont {Chen} and\ \citenamefont
  {Redner}(2005)}]{ChenRedner2005PhysRevE}%
  \BibitemOpen
  \bibfield  {author} {\bibinfo {author} {\bibfnamefont {P.}~\bibnamefont
  {Chen}} and\ \bibinfo {author} {\bibfnamefont {S.}~\bibnamefont {Redner}},\
  }\href@noop {} {\bibfield  {journal} {\bibinfo  {journal} {Phys. Rev. E}\
  }\textbf {\bibinfo {volume} {71}},\ \bibinfo {pages} {036101} (\bibinfo
  {year} {2005})}\BibitemShut {NoStop}%
\bibitem [{\citenamefont {Deffuant}\ \emph {et~al.}(2000) \citenamefont {Deffuant}, 
  \citenamefont {Neau}, \citenamefont{Amblard},and\ \citenamefont {Weisbuch}}]{DeffuantNeauAmblardWeisbuch2000AdvComplexSyst}%
  \BibitemOpen
  \bibfield  {author} {\bibinfo {author} {\bibfnamefont {G.}~\bibnamefont
  {Deffuant}}, \bibinfo {author} {\bibfnamefont {D.}~\bibnamefont
  {Neau}}, \bibinfo {author} {\bibfnamefont {F.}~\bibnamefont
  {Amblard}}, and\ \bibinfo {author} {\bibfnamefont {G.}~\bibnamefont {Weisbuch}},\ }\href@noop {}
  {\bibfield  {journal} {\bibinfo  {journal} {Adv. Complex Syst.}\ }\textbf {\bibinfo
  {volume} {3}},\ \bibinfo {pages} {87} (\bibinfo {year}{2000})}\BibitemShut {NoStop}%
\bibitem [{\citenamefont {Patriarca} and\ \citenamefont
  {{Lepp\"{a}nen}}(2004)}]{PatriarcaLeppanen2004PhysicaA}%
  \BibitemOpen
  \bibfield  {author} {\bibinfo {author} {\bibfnamefont {M.}~\bibnamefont
  {Patriarca}} and\ \bibinfo {author} {\bibfnamefont {T.}~\bibnamefont
  {{Lepp\"{a}nen}}},\ }\href@noop {} {\bibfield  {journal} {\bibinfo  {journal}
  {Physica A}\ }\textbf {\bibinfo {volume} {338}},\ \bibinfo {pages} {296}
  (\bibinfo {year} {2004})}\BibitemShut {NoStop}%
\bibitem [{\citenamefont {Patriarca} and\ \citenamefont
  {Heinsalu}(2009)}]{PatriarcaHeinsalu2009PhysicaA}%
  \BibitemOpen
  \bibfield  {author} {\bibinfo {author} {\bibfnamefont {M.}~\bibnamefont
  {Patriarca}} and\ \bibinfo {author} {\bibfnamefont {E.}~\bibnamefont
  {Heinsalu}},\ }\href@noop {} {\bibfield  {journal} {\bibinfo  {journal}
  {Physica A}\ }\textbf {\bibinfo {volume} {388}},\ \bibinfo {pages} {174}
  (\bibinfo {year} {2009})}\BibitemShut {NoStop}%
\bibitem [{\citenamefont {Mira} and\ \citenamefont
  {Paredes}(2005)}]{MiraParedes2005EurophysLett}%
  \BibitemOpen
  \bibfield  {author} {\bibinfo {author} {\bibfnamefont {J.}~\bibnamefont
  {Mira}} and\ \bibinfo {author} {\bibfnamefont {{\'{A}}.}~\bibnamefont
  {Paredes}},\ }\href@noop {} {\bibfield  {journal} {\bibinfo  {journal}
  {Europhys. Lett.}\ }\textbf {\bibinfo {volume} {69}},\ \bibinfo {pages}
  {1031} (\bibinfo {year} {2005})}\BibitemShut {NoStop}%
\bibitem [{\citenamefont {Mira}\ \emph {et~al.}(2011)\citenamefont {Mira},
  \citenamefont {Seoane},and\ \citenamefont
  {Nieto}}]{MiraSeoaneNieto2011NewJPhys}%
  \BibitemOpen
  \bibfield  {author} {\bibinfo {author} {\bibfnamefont {J.}~\bibnamefont
  {Mira}}, \bibinfo {author} {\bibfnamefont {L.~F.}\ \bibnamefont {Seoane}}, and\ 
  \bibinfo {author} {\bibfnamefont {J.~J.}\ \bibnamefont {Nieto}},\
  }\href@noop {} {\bibfield  {journal} {\bibinfo  {journal} {New J. Phys.}\
  }\textbf {\bibinfo {volume} {13}},\ \bibinfo {pages} {033007} (\bibinfo
  {year} {2011})}\BibitemShut {NoStop}%
\bibitem [{\citenamefont {{Castell\'{o}}}\ \emph {et~al.}(2006)\citenamefont
  {{Castell\'{o}}}, \citenamefont {{Egu\'{i}luz}},and\ \citenamefont {{San
  Miguel}}}]{CastelloEguiluzSanMiguel2006NewJPhys}%
  \BibitemOpen
  \bibfield  {author} {\bibinfo {author} {\bibfnamefont {X.}~\bibnamefont
  {{Castell\'{o}}}}, \bibinfo {author} {\bibfnamefont {V.~M.}\ \bibnamefont
  {{Egu\'{i}luz}}}, and\ \bibinfo {author} {\bibfnamefont {M.}~\bibnamefont
  {{San Miguel}}},\ }\href@noop {} {\bibfield  {journal} {\bibinfo  {journal}
  {New J. Phys.}\ }\textbf {\bibinfo {volume} {8}},\ \bibinfo {pages} {308}
  (\bibinfo {year} {2006})}\BibitemShut {NoStop}%
\bibitem [{\citenamefont {Vazquez}\ \emph {et~al.}(2010)\citenamefont
  {Vazquez}, \citenamefont {{Castell\'{o}}},and\ \citenamefont {{San
  Miguel}}}]{VazquezCastelloSanMiguel2010JStatMech}%
  \BibitemOpen
  \bibfield  {author} {\bibinfo {author} {\bibfnamefont {F.}~\bibnamefont
  {Vazquez}}, \bibinfo {author} {\bibfnamefont {X.}~\bibnamefont
  {{Castell\'{o}}}}, and\ \bibinfo {author} {\bibfnamefont {M.}~\bibnamefont
  {{San Miguel}}},\ }\href@noop {} {\bibfield  {journal} {\bibinfo  {journal}
  {J. Stat. Mech.}} \bibinfo {pages}{P04007} (\bibinfo {year} {2010})}\BibitemShut {NoStop}%
\bibitem [{\citenamefont {Chapel}\ \emph {et~al.}(2010)\citenamefont {Chapel},
  \citenamefont {{Castell\'{o}}}, \citenamefont {Bernard}, \citenamefont
  {Deffuant}, \citenamefont {{Egu\'{i}luz}}, \citenamefont {Martin},and\
  \citenamefont {{San
  Miguel}}}]{ChapelCastelloBernardDeffuantEguiluzMartinMiguel2010PLoSONE}%
  \BibitemOpen
  \bibfield  {author} {\bibinfo {author} {\bibfnamefont {L.}~\bibnamefont
  {Chapel}}, \bibinfo {author} {\bibfnamefont {X.}~\bibnamefont
  {{Castell\'{o}}}}, \bibinfo {author} {\bibfnamefont {C.}~\bibnamefont
  {Bernard}}, \bibinfo {author} {\bibfnamefont {G.}~\bibnamefont {Deffuant}},
  \bibinfo {author} {\bibfnamefont {V.~M.}\ \bibnamefont {{Egu\'{i}luz}}},
  \bibinfo {author} {\bibfnamefont {S.}~\bibnamefont {Martin}}, and\ \bibinfo
  {author} {\bibfnamefont {M.}~\bibnamefont {{San Miguel}}},\ }\href@noop {}
  {\bibfield  {journal} {\bibinfo  {journal} {PLOS ONE}\ }\textbf {\bibinfo
  {volume} {5}},\ \bibinfo {pages} {e8681} (\bibinfo {year}
  {2010})}\BibitemShut {NoStop}%
\bibitem [{\citenamefont {Fujie}\ \emph {et~al.}()\citenamefont {Fujie},
  \citenamefont {Aihara},and\ \citenamefont
  {Masuda}}]{FujieAiharaMasuda2012JStatPhys}%
  \BibitemOpen
  \bibfield  {author} {\bibinfo {author} {\bibfnamefont {R.}~\bibnamefont
  {Fujie}}, \bibinfo {author} {\bibfnamefont {K.}~\bibnamefont {Aihara}}, and\ 
  \bibinfo {author} {\bibfnamefont {N.}~\bibnamefont {Masuda}},\
  }\href@noop {}{\bibfield  {journal} {\bibinfo  {journal} {J. Stat. Phys.}\ }
  \textbf {\bibinfo{volume} {151}},\ \bibinfo {pages} {289} (\bibinfo {year}{2013})}\BibitemShut {NoStop}%
\bibitem [{\citenamefont {Vazquez}\ \emph {et~al.}(2008)\citenamefont
  {Vazquez}, \citenamefont {Egu\'{\i}luz},and\ \citenamefont {{San
  Miguel}}}]{Vazquez2008}%
  \BibitemOpen
  \bibfield  {author} {\bibinfo {author} {\bibfnamefont {F.}~\bibnamefont
  {Vazquez}}, \bibinfo {author} {\bibfnamefont {V.~M.}\ \bibnamefont {{Egu\'{i}luz}}}, 
  and\ \bibinfo {author} {\bibfnamefont {M.}~\bibnamefont
  {{San Miguel}}},\ }\href {\doibase 10.1103/PhysRevLett.100.108702} {\bibfield
   {journal} {\bibinfo  {journal} {Phys. Rev. Lett.}\ }\textbf {\bibinfo
  {volume} {100}},\ \bibinfo {pages} {108702} (\bibinfo {year}
  {2008})}\BibitemShut {NoStop}%
\bibitem [{\citenamefont {Kozma} and\ \citenamefont
  {Barrat}(2008)}]{Kozma2008}%
  \BibitemOpen
  \bibfield  {author} {\bibinfo {author} {\bibfnamefont {B.}~\bibnamefont
  {Kozma}} and\ \bibinfo {author} {\bibfnamefont {A.}~\bibnamefont {Barrat}},\
  }\href {\doibase 10.1103/PhysRevE.77.016102} {\bibfield  {journal} {\bibinfo
  {journal} {Phys. Rev. E}\ }\textbf {\bibinfo {volume} {77}},\ \bibinfo
  {pages} {016102} (\bibinfo {year} {2008})}\BibitemShut {NoStop}%
\bibitem [{\citenamefont {Holme} and\ \citenamefont
  {Newman}(2006)}]{Holme2006}%
  \BibitemOpen
  \bibfield  {author} {\bibinfo {author} {\bibfnamefont {P.}~\bibnamefont
  {Holme}} and\ \bibinfo {author} {\bibfnamefont {M.~E.~J.}~\bibnamefont {Newman}},\
  }\href {\doibase 10.1103/PhysRevE.74.056108} {\bibfield  {journal} {\bibinfo
  {journal} {Phys. Rev. E}\ }\textbf {\bibinfo {volume} {74}},\ \bibinfo
  {pages} {056108} (\bibinfo {year} {2006})}\BibitemShut {NoStop}%
\bibitem [{\citenamefont {Nardini}\ \emph {et~al.}(2008)\citenamefont
  {Nardini}, \citenamefont {Kozma},and\ \citenamefont
  {Barrat}}]{Nardini2008}%
  \BibitemOpen
  \bibfield  {author} {\bibinfo {author} {\bibfnamefont {C.}~\bibnamefont
  {Nardini}}, \bibinfo {author} {\bibfnamefont {B.}~\bibnamefont {Kozma}}, and\ 
  \bibinfo {author} {\bibfnamefont {A.}~\bibnamefont {Barrat}},\ }\href
  {\doibase 10.1103/PhysRevLett.100.158701} {\bibfield  {journal} {\bibinfo
  {journal} {Phys. Rev. Lett.}\ }\textbf {\bibinfo {volume} {100}},\ \bibinfo
  {pages} {158701} (\bibinfo {year} {2008})}\BibitemShut {NoStop}%
\bibitem [{\citenamefont {Masuda}\ \emph {et~al.}(2010)\citenamefont {Masuda},
  \citenamefont {Gibert},and\ \citenamefont
  {Redner}}]{MasudaGibertRedner2010PRE}%
  \BibitemOpen
  \bibfield  {author} {\bibinfo {author} {\bibfnamefont {N.}~\bibnamefont
  {Masuda}}, \bibinfo {author} {\bibfnamefont {N.}~\bibnamefont {Gibert}}, and\ 
  \bibinfo {author} {\bibfnamefont {S.}~\bibnamefont {Redner}},\
  }\href@noop {} {\bibfield  {journal} {\bibinfo  {journal} {Phys. Rev. E}\
  }\textbf {\bibinfo {volume} {82}},\ \bibinfo {pages} {010103(R)} (\bibinfo
  {year} {2010})}\BibitemShut {NoStop}%
\bibitem [{\citenamefont {Masuda} and\ \citenamefont
  {Redner}(2011)}]{MasudaRedner2011JStatMech}%
  \BibitemOpen
  \bibfield  {author} {\bibinfo {author} {\bibfnamefont {N.}~\bibnamefont
  {Masuda}} and\ \bibinfo {author} {\bibfnamefont {S.}~\bibnamefont
  {Redner}},\ }\href@noop {} {\bibfield  {journal} {\bibinfo  {journal} {J.
  Stat. Mech.}} \bibinfo {pages} {L02002} (\bibinfo {year} {2011})}\BibitemShut {NoStop}%
\bibitem [{\citenamefont {Dixit} and\ \citenamefont
  {Weibull}(2007)}]{DixitWeibull2007PNAS}%
  \BibitemOpen
  \bibfield  {author} {\bibinfo {author} {\bibfnamefont {A.~K.}\ \bibnamefont
  {Dixit}} and\ \bibinfo {author} {\bibfnamefont {J.~W.}\ \bibnamefont
  {Weibull}},\ }\href@noop {} {\bibfield  {journal} {\bibinfo  {journal} {Proc.
  Natl. Acad. Sci. U.S.A.}\ }\textbf {\bibinfo {volume} {104}},\ \bibinfo
  {pages} {7351} (\bibinfo {year} {2007})}\BibitemShut {NoStop}%
\bibitem [{\citenamefont {Zimper} and\ \citenamefont
  {Ludwig}(2009)}]{ZimperLudwig2009JRiskUncertain}%
  \BibitemOpen
  \bibfield  {author} {\bibinfo {author} {\bibfnamefont {A.}~\bibnamefont
  {Zimper}} and\ \bibinfo {author} {\bibfnamefont {A.}~\bibnamefont
  {Ludwig}},\ }\href@noop {} {\bibfield  {journal} {\bibinfo  {journal} {J.
  Risk Uncertainty}\ }\textbf {\bibinfo {volume} {39}},\ \bibinfo {pages} {181}
  (\bibinfo {year} {2009})}\BibitemShut {NoStop}%
\bibitem [{\citenamefont {Acemoglu}\ \emph {et~al.}()\citenamefont {Acemoglu},
  \citenamefont {Victor},and\ \citenamefont
  {Muhamet}}]{AcemogluVictorMuhamet2009Fragility}%
  \BibitemOpen
  \bibfield  {author} {\bibinfo {author} {\bibfnamefont {D.}~\bibnamefont
  {Acemoglu}}, \bibinfo {author} {\bibfnamefont {C.}~\bibnamefont {Victor}}, \
  and\ \bibinfo {author} {\bibfnamefont {Y.}~\bibnamefont {Muhamet}},\
  }\href@noop {} {\enquote {\bibinfo {title} {Fragility of asymptotic agreement
  under bayesian learning},}\ }
  \bibinfo {howpublished}{{http://econ-www.mit.edu/files/3795} (2009, Accessed: Febrary 1, 2013)}
  \BibitemShut {NoStop}%
\bibitem [{\citenamefont {Acemoglu} and\ \citenamefont
  {Ozdaglar}(2011)}]{AcemogluOzdaglar2011DynGamesAppl}%
  \BibitemOpen
  \bibfield  {author} {\bibinfo {author} {\bibfnamefont {D.}~\bibnamefont
  {Acemoglu}} and\ \bibinfo {author} {\bibfnamefont {A.}~\bibnamefont
  {Ozdaglar}},\ }\href@noop {} {\bibfield  {journal} {\bibinfo  {journal} {Dyn.
  Games Appl.}\ }\textbf {\bibinfo {volume} {1}},\ \bibinfo {pages} {3}
  (\bibinfo {year} {2011})}\BibitemShut {NoStop}%
\bibitem [{\citenamefont {Andreoni} and\ \citenamefont
  {Mylovanov}(2012)}]{AndreoniMylovanov2012AmEconJ-Microecon}%
  \BibitemOpen
  \bibfield  {author} {\bibinfo {author} {\bibfnamefont {J.}~\bibnamefont
  {Andreoni}} and\ \bibinfo {author} {\bibfnamefont {T.}~\bibnamefont
  {Mylovanov}},\ }\href@noop {} {\bibfield  {journal} {\bibinfo  {journal} {Am.
  Econ. J. Microecon.}\ }\textbf {\bibinfo {volume} {4}},\ \bibinfo {pages}
  {209} (\bibinfo {year} {2012})}\BibitemShut {NoStop}%
\bibitem [{\citenamefont {Binmore}(2008)}]{Binmore2008RationalDecisions}%
  \BibitemOpen
  \bibfield  {author} {\bibinfo {author} {\bibfnamefont {K.}~\bibnamefont
  {Binmore}},\ }\href@noop {} 
  {\bibinfo {title} {\textit{Rational Decisions}}}\
  (\bibinfo  {publisher} {Princeton University Press},\ \bibinfo {address}
  {Princeton, NJ},\ \bibinfo {year} {2008})\BibitemShut {NoStop}%
\bibitem [{\citenamefont {Martins}(2009)}]{Martins2009JStatMech}%
  \BibitemOpen
  \bibfield  {author} {\bibinfo {author} {\bibfnamefont {A.~C.~R.}\ \bibnamefont
  {Martins}},\ }\href@noop {} {\bibfield  {journal} {\bibinfo  {journal}
  {J. Stat. Mech.}} \bibinfo {pages}{P02017} (\bibinfo {year} {2009})}\BibitemShut {NoStop}%
\bibitem [{\citenamefont
  {Plous}(1993)}]{Plous1993ThePsychologyofJudgmentandDecisionMaking}%
  \BibitemOpen
  \bibfield  {author} {\bibinfo {author} {\bibfnamefont {S.}~\bibnamefont
  {Plous}},\ }\href@noop {} 
  {\bibinfo {title} {\textit{The Psychology of Judgment and Decision Making}}}\ 
  (\bibinfo  {publisher} {McGraw-Hill},\
  \bibinfo {address} {New York},\ \bibinfo {year} {1993})\BibitemShut
  {NoStop}%
\bibitem [{\citenamefont {Nickerson}(1998)}]{Nickerson1998RevGenPsychol}%
  \BibitemOpen
  \bibfield  {author} {\bibinfo {author} {\bibfnamefont {R.~S.}\ \bibnamefont
  {Nickerson}},\ }\href@noop {} {\bibfield  {journal} {\bibinfo  {journal}
  {Rev. Gen. Psychol.}\ }\textbf {\bibinfo {volume} {2}},\ \bibinfo {pages}
  {175} (\bibinfo {year} {1998})}\BibitemShut {NoStop}%
\bibitem [{\citenamefont {Deffuant} and\ \citenamefont{Huet}(2010)}]{DeffuantHuet2009COMPLEXITY}%
  \BibitemOpen
  \bibfield  {author} {
  \bibinfo {author} {\bibfnamefont {G.}~\bibnamefont{Deffuant}} and\ 
  \bibinfo {author} {\bibfnamefont {S.}~\bibnamefont{Huet}},\ }\href@noop {} {\bibfield  {journal} {\bibinfo  {journal} {Complexity}\ }\textbf {\bibinfo {volume} {15}}(5),\ \bibinfo {pages} {25} (\bibinfo {year} {2010})}\BibitemShut {NoStop}%
\bibitem [{\citenamefont {Rabin} and\ \citenamefont
  {Schrag}(1999)}]{RabinSchrag1999QJECON}%
  \BibitemOpen
  \bibfield  {author} {\bibinfo {author} {\bibfnamefont {M.}~\bibnamefont
  {Rabin}} and\ \bibinfo {author} {\bibfnamefont {J.~L.}\ \bibnamefont
  {Schrag}},\ }\href@noop {} {\bibfield  {journal} {\bibinfo  {journal} {Q. J.
  Econ.}\ }\textbf {\bibinfo {volume} {114}},\ \bibinfo {pages} {37} (\bibinfo
  {year} {1999})}\BibitemShut {NoStop}%
\bibitem [{\citenamefont {{Orl\'{e}an}}(1995)}]{Orlean1995JEconBehavOrgan}%
  \BibitemOpen
  \bibfield  {author} {\bibinfo {author} {\bibfnamefont {A.}~\bibnamefont
  {{Orl\'{e}an}}},\ }\href@noop {} {\bibfield  {journal} {\bibinfo  {journal}
  {J. Econ. Behav. Organ.}\ }\textbf {\bibinfo {volume} {28}},\ \bibinfo
  {pages} {257} (\bibinfo {year} {1995})}\BibitemShut {NoStop}%
\bibitem [{\citenamefont {{P\'{e}rez-Escudero}} and\ \citenamefont
  {de~Polavieja}(2011)}]{Perez-EscuderodePolavieja2011PLoSComputBiol}%
  \BibitemOpen
  \bibfield  {author} {\bibinfo {author} {\bibfnamefont {A.}~\bibnamefont
  {{P\'{e}rez-Escudero}}} and\ \bibinfo {author} {\bibfnamefont {G.~G.}\
  \bibnamefont {de~Polavieja}},\ }\href@noop {} {\bibfield  {journal} {\bibinfo
   {journal} {PLOS Comput. Biol.}\ }\textbf {\bibinfo {volume} {7}},\ \bibinfo
  {pages} {e1002282} (\bibinfo {year} {2011})}\BibitemShut {NoStop}%
\bibitem [{\citenamefont {Pacheco}\ \emph
  {et~al.}(2009{\natexlab{a}})\citenamefont {Pacheco}, \citenamefont {Santos},
  \citenamefont {Souza},and\ \citenamefont
  {Skyrms}}]{PachecoSantosSouzaSkyrms2009ProcRSocB}%
  \BibitemOpen
  \bibfield  {author} {\bibinfo {author} {\bibfnamefont {J.~M.}\ \bibnamefont
  {Pacheco}}, \bibinfo {author} {\bibfnamefont {F.~C.}\ \bibnamefont {Santos}},
  \bibinfo {author} {\bibfnamefont {M.~O.}\ \bibnamefont {Souza}}, and\
  \bibinfo {author} {\bibfnamefont {B.}~\bibnamefont {Skyrms}},\ }\href@noop {}
  {\bibfield  {journal} {\bibinfo  {journal} {Proc. R. Soc. B}\ }\textbf
  {\bibinfo {volume} {276}},\ \bibinfo {pages} {315} (\bibinfo {year}
  {2009}{\natexlab{a}})}\BibitemShut {NoStop}%
\bibitem [{\citenamefont {Pacheco}\ \emph
  {et~al.}(2009{\natexlab{b}})\citenamefont {Pacheco}, \citenamefont
  {Pinheiro},and\ \citenamefont
  {Santos}}]{PachecoPinheiroSantos2009PLoSComputBiol}%
  \BibitemOpen
  \bibfield  {author} {\bibinfo {author} {\bibfnamefont {J.~M.}\ \bibnamefont
  {Pacheco}}, \bibinfo {author} {\bibfnamefont {F.~L.}\ \bibnamefont
  {Pinheiro}}, and\ \bibinfo {author} {\bibfnamefont {F.~C.}\ \bibnamefont
  {Santos}},\ }\href@noop {} {\bibfield  {journal} {\bibinfo  {journal} {PLOS.
  Comput. Biol.}\ }\textbf {\bibinfo {volume} {5}},\ \bibinfo {pages}
  {e1000596} (\bibinfo {year} {2009}{\natexlab{b}})}\BibitemShut {NoStop}%
\bibitem [{\citenamefont {Baliga}\ \emph {et~al.}()\citenamefont {Baliga},
  \citenamefont {Hanany},and\ \citenamefont
  {Klibanoff}}]{BaligaHananyKlibanoff2011Polarization}%
  \BibitemOpen
  \bibfield  {author} {\bibinfo {author} {\bibfnamefont {S.}~\bibnamefont
  {Baliga}}, \bibinfo {author} {\bibfnamefont {E.}~\bibnamefont {Hanany}}, and\ 
  \bibinfo {author} {\bibfnamefont {P.}~\bibnamefont {Klibanoff}},\
  }\href@noop {} {\enquote {\bibinfo {title} {Polarization and ambiguity},}\
  }
  \bibinfo {note} {Am. Econ. Rev. (to be published)}\BibitemShut{NoStop}%
\bibitem [{\citenamefont {Masuda}(2009)}]{Masuda2009JTheorBiol}%
  \BibitemOpen
  \bibfield  {author} {\bibinfo {author} {\bibfnamefont {N.}~\bibnamefont
  {Masuda}},\ }\href@noop {} {\bibfield  {journal} {\bibinfo  {journal} {J.
  Theor. Biol.}\ }\textbf {\bibinfo {volume} {258}},\ \bibinfo {pages} {323}
  (\bibinfo {year} {2009})}\BibitemShut {NoStop}%
\bibitem [{\citenamefont {Doll}\ \emph {et~al.}(2011)\citenamefont {Doll},
  \citenamefont {Hutchison},and\ \citenamefont
  {Frank}}]{DollHutchisonFrank2011JNeurosci}%
  \BibitemOpen
  \bibfield  {author} {\bibinfo {author} {\bibfnamefont {B.~B.}\ \bibnamefont
  {Doll}}, \bibinfo {author} {\bibfnamefont {K.~E.}\ \bibnamefont {Hutchison}},
  and\ \bibinfo {author} {\bibfnamefont {M.~J.}\ \bibnamefont {Frank}},\
  }\href@noop {} {\bibfield  {journal} {\bibinfo  {journal} {J. Neurosci.}\
  }\textbf {\bibinfo {volume} {31}},\ \bibinfo {pages} {6188} (\bibinfo {year}
  {2011})}\BibitemShut {NoStop}%
\end{thebibliography}

\providecommand{\noopsort}[1]{}\providecommand{\singleletter}[1]{#1}%

\end{document}